\documentclass[aps,pre,psfig,onecolumn,showpacs,superscriptaddress,preprint,floatfix]{revtex4-1}
\usepackage{amsmath}
\usepackage{mathtools}
\usepackage{amssymb}
\usepackage{color}
\usepackage{graphics}
\usepackage{epsfig}
\usepackage[normalem]{ulem} 
\usepackage{cancel}
\usepackage[mathlines]{lineno}
\graphicspath{{Figures/}}
\usepackage{svg}
\usepackage{xtab,afterpage,longtable}
\usepackage{booktabs}   
\usepackage{ltablex}
\usepackage{enumitem}
\relax
\usepackage{float}
\usepackage{orcidlink}
\usepackage{hyperref}
\hypersetup{colorlinks=true,linkcolor=blue,urlcolor=blue,citecolor=blue}

\begin{document}
\title{Nonlinear Mixing of Waves in a Yukawa One Component Plasma}
\author{Ajaz Mir\orcidlink{0000-0001-5540-4967}}
\email{ajazmir.physics@gmail.com}
\affiliation{Institute for Plasma Research, Gandhinagar 382428, Gujarat, India} 
\author{Farida Batool\orcidlink{0000-0002-8760-8005}}
\author{Sanat Tiwari\orcidlink{0000-0002-6346-9952}}
\affiliation{Indian Institute of Technology Jammu, Jammu 181221, J\&K, India}
\author{Abhijit Sen\orcidlink{0000-0001-9878-4330}}
\affiliation{Institute for Plasma Research, Gandhinagar 382428, Gujarat, India} 
\affiliation{Homi Bhabha National Institute, Anushaktinagar, Mumbai 400094, India}
\date{\today}
\begin{abstract}
The phenomenon of nonlinear wave mixing is investigated in a Yukawa one-component plasma using two-dimensional classical Langevin molecular dynamics simulations. The wave spectrum indicates that nonlinear interactions between the excited modes are primarily governed by a three-wave mixing mechanism, as confirmed by bispectral analysis. In particular, the mixing characteristics observed in the simulations closely resemble those reported in previous numerical studies of the forced Korteweg-de Vries (fKdV) model [ \href{https://doi.org/10.1063/5.0077638}{\textcolor{blue}{\text{Phys. Plasmas {\bf 29}, 032303 (2022)}}}]. This similarity further validates the applicability of the fKdV fluid model in capturing the weakly nonlinear dynamics of dusty plasmas with reasonable accuracy.  
\end{abstract}
\maketitle
\section{Introduction}
\label{intro}
\paragraph*{} 
Nonlinear mixing (NLM) describes the generation of a cascade of waves (modes) resulting from the interaction of two or more primary waves (modes) in a nonlinear medium~\cite{Ajaz_POP_2020, Nosenko_PRL_2004, Misoguti_PRA_2005, Zheng_APX_2021}. It occurs in a variety of physical systems and plays a key role in many practical applications, e.g., in optical media~\cite{Liu_APL_2020, Buono_OPL_2018}, in plasmas~\cite{Nosenko_PRL_2004, Papen_JPP_1989}, in neutral fluids~\cite{Li_PRE_2021}, and in material science~\cite{Lewandowski_PRB_2016, Cox_ACS_2015}. 
Some typical examples of NLM include the mixing of vortices in a Kerr-like nonlinear medium~\cite{Lenzini_PRA_2011}, generation of short wavelength light due to laser interactions~\cite{Misoguti_PRA_2005}, and nonlinear mixing in meta-materials~\cite{Huang_APL_2011, Shadrivov_APL_2008}. Recently, we demonstrated the success of a continuum fluid-based forced Korteweg-de Vries (fKdV) model equation ~\cite{Ajaz_POP_2020} in capturing the characteristics of NLM in a dusty plasma medium. 
The dominant NLM process was found to be a three-wave mixing process as confirmed by a bispectral analysis~\cite{Ajaz_POP_2022}.
\paragraph*{}
\textcolor{black}{The primary motivation of our present work is to investigate the phenomenon of nonlinear wave mixing at a kinetic level by carrying out detailed molecular dynamic simulations. In doing so we also establish a strong correlation with past results that were obtained using the fKdV model equation~\cite{Ajaz_POP_2022}. As is well known, the Korteweg-de Vries (KdV) model is derived from model fluid equations under restrictive assumptions and its applicability to various experimental conditions has therefore been somewhat debatable. Our present work provides a firmer physical basis for the applicability of the KdV model by demonstrating that the simulation results align remarkably well with the analytic predictions of the KdV model. This is a novel result and a significant advancement over past work that helps to extend the utility of the KdV model over a broad range of nonlinear phenomena by bridging the gap between theory, simulation, and real-world plasma dynamics.}
\paragraph*{}
\textcolor{black}{
In a typical electron-ion plasma, when the focus is on ion dynamics, the electrons primarily serve as a background medium that shields ion-ion interactions. In such cases, the ions can be effectively treated as a one-component plasma (OCP), interacting through a shielded potential without explicitly modeling the electrons. Similarly, in dusty plasmas, the combined presence of electrons and ions provides the background shielding for dust particles, which evolve on much slower timescales. This enables the dust component to be modeled as an OCP as well, where dust particles interact via a screened potential, again without explicitly resolving the electrons and ions. Hence, our simulations, although for dusty plasma, also represent generic one-component plasmas across different time and length scales. In both of these contexts, the screening effect leads to a pairwise interaction between like-charged particles that takes the form of a screened Coulomb or Yukawa potential. In general, the Yukawa one-component plasma (YOCP) model is a framework used to describe plasmas consisting of one component of interest, while the remaining components act as a neutralizing and screening background.
The YOCP model applies to various physical systems, such as dusty plasmas~\cite{Oxtoby_PRL_2013, Lin_PRE_2019, Marciante_PRL_2017}, ultra-cold plasma~\cite{Langin_PRE_2016}, non-neutral plasma~\cite{Dubin_RMP_1999}, and warm dense matter~\cite{Kahlert_PRE_2020, Bonitz_POP_2020}, all of which exhibit behavior characteristic of Yukawa plasma systems. 
}
\paragraph*{}
The YOCP model provides a good representation of the dynamics of charged dust particles in a dusty plasma and has been used in the past to investigate nonlinear structures like solitons~\cite{Tiwari_POP_2015, Sandeep_POP_2017}. 
The model is characterized by two physical parameters - \textcolor{black}{the screening parameter $\kappa = a / \lambda_D$} and the Coulomb coupling strength $\Gamma = Q^2/(4\pi \epsilon_0 a k_B T)$. Here \( a = \sqrt{1/ (\pi n)} \) is the average interparticle distance, \( n \) the areal number density of the particles, \( \lambda_D \) is the Debye length, and \( T \) is the kinetic temperature of the particles. The time evolution dynamics of OCP in 2D is characterized by the inverse of its characteristic frequency \( \omega_{pd} = \sqrt{nQ^2 / (2a \epsilon_0 m )} \). Here, \( Q \) and \( m \) are the charge and mass of the particles. Molecular dynamics simulations serve as a robust tool to investigate nonlinear phenomena at the particle level and offer a means of testing the validity/limitations of the fluid approach~\cite{Tiwari_POP_2015, Sandeep_POP_2017}.
\paragraph*{}
\textcolor{black}{
The KdV model describes the weakly nonlinear dynamics of dusty plasmas, a typical Yukawa system. In general, the KdV model applies to a wide range of inter-particle potentials, including those in 1D and 2D lattice structures~\cite{Avinash_PRE_2003} and has been used to successfully explain experimental observations of the propagation of nonlinear compressional waves in 2D Yukawa lattices~\cite{Nosenko_PRL_2002}. In the presence of a driving source, a generalization of the model in the form of the fKdV equation, has served as a good paradigmatic model for nonlinear phenomena observed in various dusty plasma experiments~\cite{Ajaz_POP_2020, Surabhi_PRE_2016, Krishan_POP_2022, Garima_PRE_2021}.}
\paragraph*{}
The manuscript is organized as follows.
The details of the Langevin MD simulations for the YOCP model system are described in section~\ref{LMD_SIM}. 
The fKdV model adopted to explain the NLM in 2D Yukawa systems is described in section~\ref{fKdV_model}. 
The characterization techniques used to quantify the nature of nonlinear interactions and the strength of dominant excited modes are explained based on power spectrum and bispectrum analysis in Section~\ref{Bspectral_analysis_method}.
A comparison is made between the mixing characteristics obtained from the simulations and those of the fKdV model, and their remarkable equivalence and some of the physical and statistical differences are pointed out in section~\ref{NLM_MD_fKdV}. 
A summary and some concluding remarks are provided in section~\ref{sum_con}.
\section{Langevin Molecular Dynamics Simulations}
\label{LMD_SIM}
We have performed 2D classical Langevin MD simulations~\cite{Nosenko_PRL_2004, Vaulina_PRL_2009, Ott_PRL_2009, Feng_PRE_2010} using the open source Large-scale Atomic/Molecular Massively Parallel Simulator (LAMMPS)~\cite{LAMMPS_2022} code to investigate the characteristics of NLM in a dusty plasma system modeled as a YOCP system. The dynamics of massive charged dust particles is governed by classical physics based on Newton's equations of motion.
The simulations take into account dissipation due to frictional gas damping by using the Langevin dynamics, which models collisions with gas atoms. To study NLM, the system is subjected to a localized external perturbation force $F_L$ that excites the primary waves. The particles are confined by an external confinement force $F_C$ at the x-boundaries. The Langevin equation of motion, which mimics the dynamics of each particle in a frictional YOCP along with the external excitation and confinement forces, is given by
\begin{equation}
    m \ddot{\bf{r}}  = - \nabla  \phi^{Y} -   m \bar{\nu} \dot{\bf{r}} + \xi (t) + F_{L} + F_{C}.
    \label{eqn_EOM}
\end{equation}
The particle trajectories $\textbf{r} (t)$ are generated for all particles by integrating Eq.~\ref{eqn_EOM} using standard Langevin dynamics~\cite{Allen_OUP_2017}. The \textcolor{black}{pair-wise} repulsive Yukawa/Debye-H\"{u}ckel~\cite{Yukawa_PPMSP_1935, Debye_ZP_1923, Rauoof_SR_2022} interaction potential between \textcolor{black}{each pair of particles each of mass $m$ and charge $Q$, which are separated by a distance $r_{ij}$} is given by~\cite{Kalman_PRL_2000}
\begin{equation}
    \phi^{Y} (r_{ij}) = Q^2(4 \pi \epsilon_0 r_{ij})^{-1} \exp(-\kappa r_{ij}) .
    \label{eqn_Yukawa}
\end{equation}
The system is confined on the left and right boundaries by repulsive forces that have a Gaussian profile of the form~\cite{Qiu_POP_2021, Lin_PRE_2019}
\begin{eqnarray}
    F_C^{Left}   = A_{C}^{Left} \exp(-(x  - x_C^{Left})^2/\sigma^2); \label{Eqn_CNF_L}
    \\   
    A_{C}^{Left} = 100\ ma\omega_{pd}^2, x_C^{Left}  = 0, \sigma = 2 a. \nonumber
    \\
    F_C^{Right}  = A_{C}^{Right}  \exp(-(x  - x_C^{Right} )^2/\sigma^2); \label{Eqn_CNF_R}
    \\
    A_{C}^{Right}  = 100\ ma\omega_{pd}^2, x_C^{Right}   = L_x, \sigma = 2 a.  \nonumber
\end{eqnarray}
Two primary waves of frequency $f_1$ and $f_2$ are excited in the system by localized perturbation forces $F_L$ that are assumed to have Gaussian profiles to mimic radiation pressure forces exerted by an external laser in an experimental situation~\cite{Nunomura_PRE_2003}.
\begin{eqnarray}
    F_L^{f_1}   = A_{L}^{f_1} \exp(-(x  - x_L^{f_1})^2/\sigma^2)\left[1 - \cos(2\pi f_1 t) \right];  \label{Eqn_WEF_L}
    \\ 
   A_{L}^{f_1} = 0.25\ ma\omega_{pd}^2,  x_L^{f_1}  = 0.2 L_x, \sigma = 2 a. \nonumber
    \\
    F_L^{f_2}  = A_{L}^{f_2} \exp(-(x  - x_L^{f_2})^2/\sigma^2)\left[1 - \cos(2\pi f_2 t) \right]; \label{Eqn_WEF_R}
    \\
   A_{L}^{f_2} = 0.25\ ma\omega_{pd}^2, x_L^{f_2}  = 0.8 L_x, \sigma = 2 a. \nonumber
\end{eqnarray}
\begin{table}[!htb]
\caption{Particle parameters used for the Langevin simulations~\cite{Nosenko_PRL_2002, Morfill_PRL_1994, Marciante_PRL_2017, Pieper_PRL_1996}.}
\begin{center}
\def~{\hphantom{0}}
{\renewcommand{\arraystretch}{1.5}
\begin{tabular}{|p{8.0cm}|p{7.0cm}|}
\hline
\textbf{Particle parameter}        &  \textbf{Numerical value} 
\\ \hline
\textbf{Dimensions of simulation box, $Lx \times Ly$}     & $L_x = 10 L_y = 560.5a$         
\\  \hline
\textbf{No. of particles, $N$}     & $10^4$                             
\\ \hline
\textbf{Number density,   $n$}     & $2.923 \times 10^6$ m$^{-2}$        
\\ \hline
\textbf{Charge of \textcolor{black}{dust} particle,  $Q$}  & 
\shortstack{\\ \textbf{$-16\times 10^3 |e|$ }  \\ $|e|$ = Magnitude of electron charge}   
\\ \hline
\textbf{Mass of \textcolor{black}{dust} particle,  $m$}    & $6.9 \times  10^{-13}$  kg 
\\ \hline
\textbf{Dust plasma frequency,  $\omega_{pd}$} $\mid$ $f_{pd}$    & $69.18~$rad/Sec $\mid$ $10.98$ Hz
\\ \hline
\textbf{Forcing frequencies, $f_{1}, f_2$}    & $0.70$ Hz, $1.70$ Hz
\\ \hline
\end{tabular}
}
\label{Tab_1}
\end{center}
\end{table}
The term $m \bar{\nu} \dot{\bf{r}} $ is the neutral drag force with drag coefficient \textcolor{black}{ $\nu = \bar{\nu} / \omega_{pd}$ } and $\xi (t)$ is the random force due to random kicking of neutral gas molecules~\cite{Feng_PRE_2008, Ott_PRL_2009}. \textcolor{black}{In other words, $\xi (t)$ is the contribution from the collision term, between the dust particle and the background neutral gas.} 
We assume that the random force $\xi (t)$, has a Gaussian distribution with zero mean \textit{i.e.,} $\left\langle \xi (t) \right\rangle = 0$. In the Langevin simulation, the heating quantified by temperature T and the friction quantified by $\bar{\nu}$ are explicitly coupled by the fluctuation-dissipation theorem. 
\textcolor{black}{The fluctuation-dissipation theorem couples the frictional drag and the random Langevin kick, which together constitute a heat bath.}
According to the theorem, the magnitude of random force, characterized by the width of the force, helps to achieve the desired temperature T~\cite{Gunsteren_MP_1982, Kubo_RPP_1966}
\begin{equation}
    \left\langle \xi_{i \alpha} (0) \xi_{j \beta} (t) \right\rangle =  2 k_B T m \Bar{\nu} \delta (t) \delta_{ij} \delta_{\alpha \beta},
    \label{eqn_FDT}
\end{equation}
where the Dirac delta function $\delta (t)$ indicates the localized nature of the random force $\xi (t)$ in time. $\delta_{ij}$ and $\delta_{\alpha \beta}$ are Kronecker delta symbols with $i, j \in \{1, ..., N\}$ being the particle indices and $\alpha, \beta \in \{x,y\}$ denoting the space coordinates. $ \left\langle \xi_{i \alpha} (0) \xi_{j \beta} (t) \right\rangle$ is the standard deviation of the Gaussian white noise $\xi (t)$.
\begin{figure*}
    \centering
    \includegraphics[width=\textwidth]{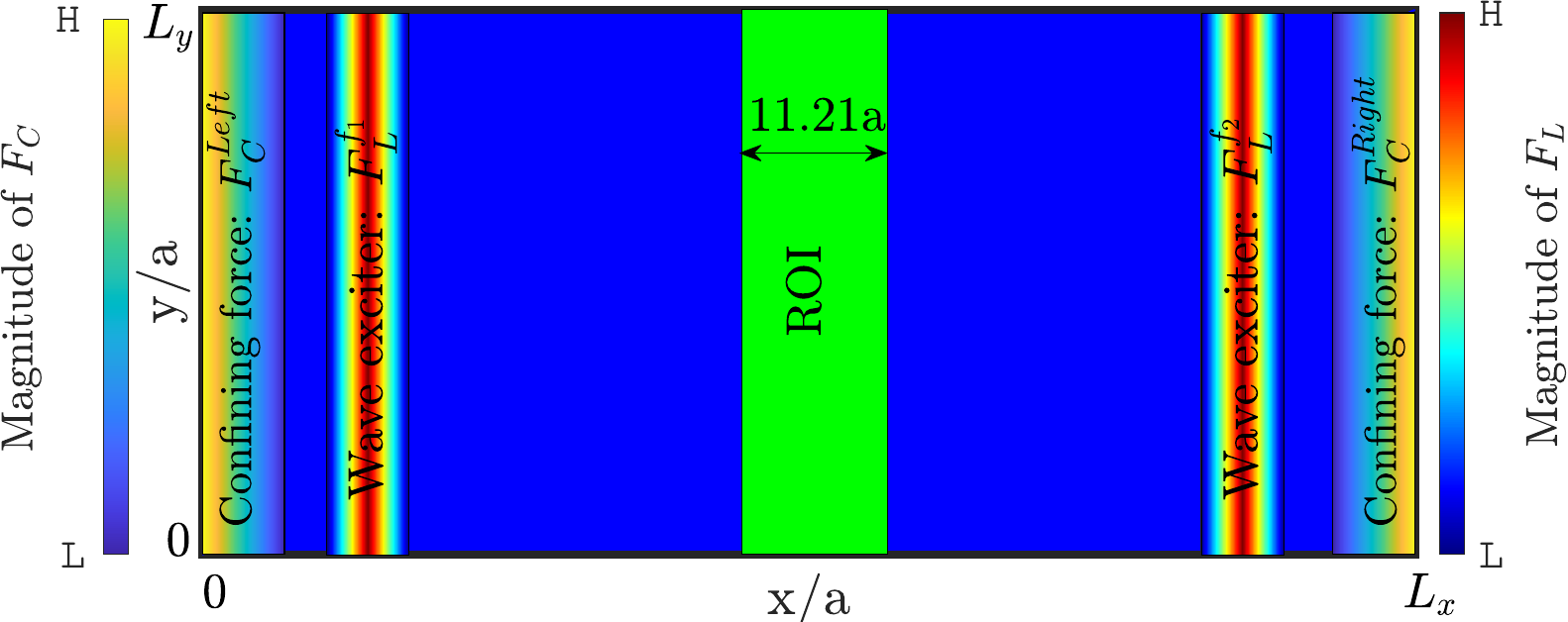}
    \caption{
    A \textcolor{black}{schematic} to demonstrate the NLM using the Langevin simulation in LAMMPS. The system is bounded at $x=0$ and $x=L_x$ by a confining force $F_C$. Two waves are excited by laser forces $F_L$, each with a different frequency. Both the $F_C$ and $F_L$ are Gaussian in nature. Periodic boundary conditions are imposed along the y-direction. The region-of-interest (ROI) used to collect time series is the green-coloured region at the centre of the rectangular ($L_x = 10 L_y$) simulation box.
    }
    \label{Fig_1}
\end{figure*}
\paragraph*{}
To carry out Langevin MD simulations using LAMMPS, we created a rectangular box ($L_x = 10 L_y$) shown in Fig.~\ref{Fig_1}, with periodic boundary conditions in the $y$ direction and reflecting boundary conditions in the $x$ direction \textcolor{black}{using the confining forces given by Eqns.~\eqref{Eqn_CNF_L}, and \eqref{Eqn_CNF_R}}. To begin with, we distribute the point-charged particles homogeneously within the box with the particle parameters given in Table~\ref{Tab_1}. Furthermore, we assigned random velocities to the particles corresponding to a temperature T. 
\textcolor{black}{
To study the dynamics of the system at a particular coupling strength $\Gamma = 100$ ($\propto$ 1/T) and $\kappa = 0.1$, the system is first equilibrated using the NVT (canonical ensemble) condition, in which the number of particles (N), volume (V), and temperature (T) are kept constant. The NVT simulations are performed such that the system exchanges heat with a heat bath to maintain a constant temperature. We equilibrate it at a specific temperature, corresponding to a given coupling strength that defines the system's state. The Nosé–Hoover thermostat~\cite{Nose_MP_1984, Hoover_PRA_1985} regulates the temperature, and we found 400 $\omega_{pd}^{-1}$ to be a reasonable time for the system to reach equilibrium around this temperature. After achieving proper equilibration, the NVT simulation is switched to the NVE (micro-canonical ensemble) condition, where the number of particles (N), volume (V), and total energy (E) are conserved. In this ensemble, the system is isolated, and there is no energy exchange with the surroundings. The primary objective of the NVE simulation is to verify that no additional free energy is present in the system; as such, excess energy could lead to destabilization and make the system unsuitable for further investigation. We evaluate the system's stability by monitoring energy fluctuations, which are within an acceptable value of $10^{-2}\%$, indicating that the system is well-prepared for subsequent studies. We also validated the system by calculating the radial distribution function and comparing it with the available results in the literature before launching the NLM studies. 
} 
\section{The forced Korteweg-de Vries model for driven dusty plasma}
\label{fKdV_model}
The theoretical framework of our approach is grounded in two key premises. First, the KdV model provides an accurate description of the evolution of weakly nonlinear and weakly dispersive compressional waves in a dusty plasma medium. Second, the fKdV equation is capable of modeling the dynamics of these nonlinear acoustic modes in the presence of an external driving force. Both premises are well-supported by prior theoretical and experimental studies. Specifically, the KdV model has been shown to effectively describe the nonlinear evolution of dust lattice waves (DLWs) and dust acoustic waves (DAWs) in dusty plasma systems~\cite{Farokhi_PRL_1999, Bandyopadhyay_PRL_2008}. The exact cnoidal wave solutions of the KdV equation govern the nonlinear evolution of DAWs in dusty plasma experiments~\cite{Liu_POP_2018}.
Furthermore, the fKdV model has been successfully applied to describe driven nonlinear ion-acoustic waves within the fluid framework of plasmas~\cite{Sen_ASR_2015}. It has also been shown to capture the dynamics of driven DAWs, including wave trapping phenomena~\cite{Krishan_POP_2022}. In particular, the fKdV model has been instrumental in interpreting the excitation of precursor and pinned dust acoustic solitons in laboratory dusty plasma experiments~\cite{Surabhi_PRE_2016, Garima_PRE_2021}. Therefore, it is reasonable to expect that the fKdV model will also be effective in describing the dynamics of driven DLWs in a 2D Yukawa lattice system subjected to external forcing~\cite{Avinash_PRE_2003}.
\paragraph*{}
\textcolor{black}{In our previous work on NLM based on the fKdV, we were able to show a good agreement between the analytical solutions of the fKdV with the observations from the mixing of DLWs in laboratory settings~\cite{Nosenko_PRL_2004}. The generalized mathematical form of the fKdV equation is given by~\cite{Ajaz_POP_2020, Salas_NLA_2011}}
\begin{equation}
\label{fKdV_eqn}
    \frac{\partial n (x,t)}{\partial t} + \alpha n (x,t) \frac{\partial n(x,t)}{\partial x} + \beta  \frac{\partial^3 n(x,t)}{\partial x^3} = F_s (t) .
\end{equation}
\textcolor{black}{The strength of nonlinearity and dispersion in the fKdV model is controlled by the coefficients $\alpha$ and $\beta$, respectively, which depend on the ambient plasma parameters~\cite{Cousens_PRE_2012, Liu_POP_2018, Sen_ASR_2015}. $F_s(t)$ is a time-dependent external source, which can be an external laser source or any fluctuating electrostatic potential source.} The NLM observed in the fKdV model was obtained through the semi-analytical solution of fKdV given by~\cite{Ajaz_POP_2022, Salas_NLA_2011}
\begin{eqnarray}
\label{fKdV_soln}
n(x, t) =\ 
\phi(t)
+\ \mu \ cn^2 \bigg[
\sqrt{\frac{ \alpha \mu}{4\beta (2\kappa_{cn} + \kappa_{cn}^2 )}} \zeta(x,t) ;\ \kappa_{cn}  
\bigg] 
\\
\zeta(x,t) = \left( x  - \frac{\kappa_{cn} + \kappa_{cn}^2 - 1}{2\kappa_{cn} + \kappa_{cn}^2} \alpha \mu t - \alpha \psi(t)\right) \nonumber
\\
\phi(t) =\ \int  F_s(t)\ dt ,\ 
\hspace{0.3cm} 
\hspace{0.3cm} 
\psi(t) =\ \int \phi(t)\ dt ,\
\nonumber
\hspace{0.3cm} 
F_s (t) =  A_s\ cn^2[2 K(\kappa_{cn}^s) f_2 t; \kappa_{cn}^s],
\nonumber
\end{eqnarray}
where $cn$ is a Jacobi elliptic function, \textcolor{black}{$\mu$ is the amplitude of the DLW, and $\kappa_{cn}$ is the elliptic parameter. The $\kappa_{cn}$ quantifies the nonlinearity of the wave, which is reflected in the shape of the wave. If $\kappa_{cn}=0$, the wave is of a linear (sinusoidal) nature, while if $\kappa_{cn}=1$, the wave is of a nonlinear (cnoidal) nature. Also, $\kappa_{cn}^s$ is again the elliptic parameter corresponding to the external source $F_s(t)$.}
The spatial wavelength $\lambda$ and the frequency $f_1$ \textcolor{black}{of DLW}, obtained from the nonlinear solution of the KdV equation (\textit{i.e.,} Eq.~\eqref{fKdV_eqn} with $A_s = 0$ ) are given by~\cite{Ajaz_POP_2020}
\begin{equation}\label{KdV_wavelength}
    \lambda = 4 K(\kappa_{cn}) \sqrt{\frac{\beta(2\kappa_{cn} + \kappa_{cn}^2)}{\alpha \mu}} 
\end{equation}
\begin{equation}\label{KdV_freq}
    f_1 = \frac{\beta (\kappa_{cn} + \kappa_{cn}^2 -1)}{4 K(\kappa_{cn})} \left( \frac{\alpha \mu}{\beta (2\kappa_{cn} + \kappa_{cn}^2)} \right)^{3/2}.
\end{equation}
\textcolor{black}{Here, $K(\kappa_{cn})$ is the complete elliptical integral of the first kind.}
The parameters used in the solution of the fKdV \textit{i.e.,} Eq.~\eqref{fKdV_soln} for which NLM is observed are $\alpha = \beta = 1;\ \mu = 18.5;\ \kappa_{cn} = \kappa_{cn}^s =  0.98$, and $A_s = \mu$. The frequency of the natural mode of the KdV corresponding to these parameters is $f_1=0.7$ Hz~\cite{Ajaz_POP_2020}.
\section{Characterization Techniques: Power spectrum and Bispectrum }
\label{Bspectral_analysis_method}
\begin{figure}
    \centering
    \includegraphics[width=\linewidth]{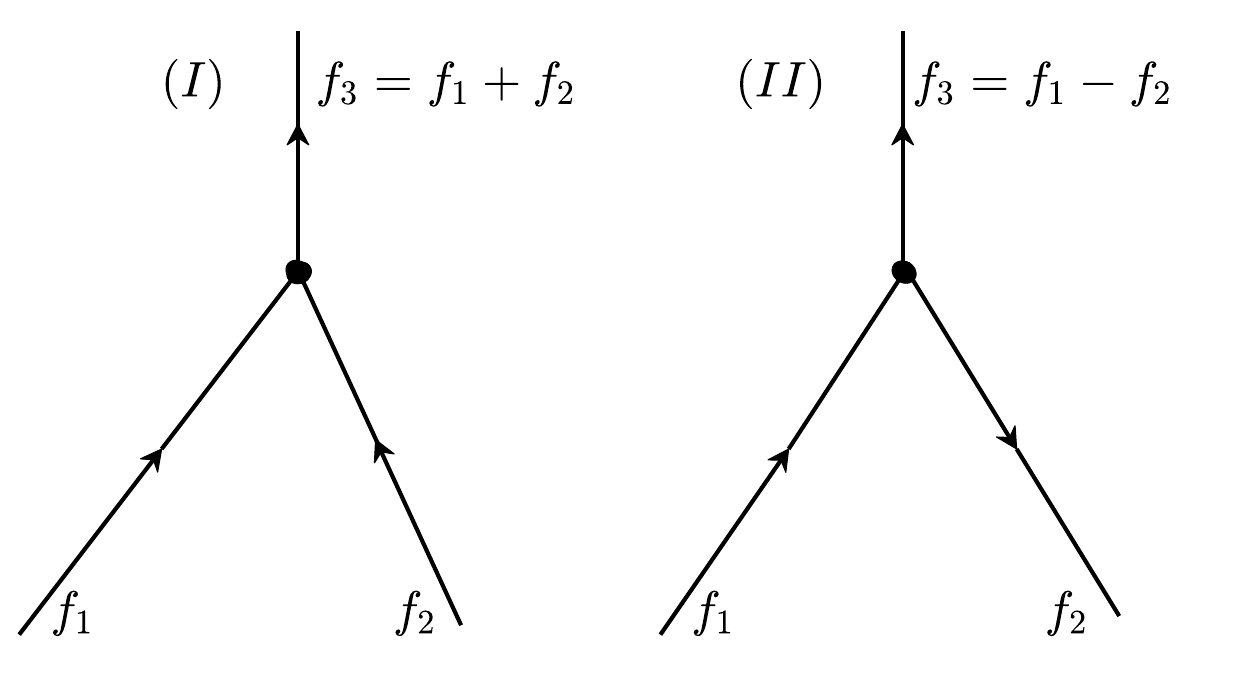}
    \caption{
    \textcolor{black}{ A schematic to illustrate a typical three-wave interaction:} (I) generic sum-interaction, (II) generic difference-interaction. If $f_1 = f_2$, (I) and (II) represent the harmonic generated by sum-self-interaction and deformation generated by difference-self-interaction, respectively.
    }
    \label{Fig_2}
\end{figure}

The power spectrum and bispectrum are the two- and three-point correlation functions in Fourier space, respectively. These are statistical tools used to extract information from the time series of a dynamical quantity. Traditionally, only the power spectrum is used to analyze the information in the continuous spectrum of propagating waves in one dimension because it is simpler to measure and model than higher-order statistics. However, the power spectrum information alone is not sufficient to establish the existence of a nonlinear origin of the wave-wave coupling event. This is because the power spectrum does not retain the phase coupling information of the nonlinear interactions.  
To establish the physical origin of three-wave coupling, a more precise tool is a bispectral analysis that looks at the triple-correlation of the time series and provides information that is supplementary to the power spectrum.  The bispectrum captures the interaction of nonlinear signals and captures phase information. A significant amount of correlation is obtained for a frequency triad $F(f_1)$, $F(f_2)$, and $F(f_1+f_2)$ when they are formed by a coherent phase coupling mechanism (where F is the discrete Fourier component at a given frequency). There will be no correlation among frequencies if they are spontaneously excited \textit{i.e.,} no coherent phase coupling is involved between the modes of nonlinear interactions.
\paragraph*{}
\textcolor{black}{As pointed out earlier in the paper, the power spectrum is built in real value, therefore, it does not carry phase information.} However, the bispectrum is capable of detecting quadratic phase coupling by correlating the signals involved in nonlinear interaction. The bispectrum is interpreted as the triple-point correlation, which reflects the energy budget of interactions between energy-supplying and energy-receiving modes of waves involved in coherent nonlinear interaction.  The turbulent cascade, which describes the transfer of energy from large \textcolor{black}{(small wave-vector \textbf{k})} to small \textcolor{black}{(large wave-vector \textbf{k})} scales of motion, is probably the most prominent consequence of triadic interactions~\cite{Monsalve_PRL_2020}. These triadic interactions are the fundamental mechanism of energy transfer in nonlinear fluid flows, that manifest themselves in Fourier space as triplets of three wavevectors, $(\textbf{k}_1, \textbf{k}_2, \textbf{k}_3)$, or frequencies, $(f_1,f_2, f_3)$ that sum to zero.
\begin{eqnarray}
    \textbf{k}_1 \pm \textbf{k}_2 \pm \textbf{k}_3 = \textbf{0};
    \hspace{1cm}
    f_1 \pm f_2 \pm f_3 = 0
\end{eqnarray}
The zero-sum condition implies that triads form triangles in the wavevector and frequency domain. An easy way to conceptually visualize these three-wave interactions is presented in Fig.~\ref{Fig_2}.
\paragraph*{}
Bispectral analysis has been extensively used in investigating the coherent nonlinear interactions in plasmas, biomedical and engineering fields~\cite{Siu_IEEE_2008, Ajaz_POP_2022, Nosenko_PRE_2006, Rao_Gabr_SV_1984}. If there is a coherent nonlinear interaction between three oscillations at frequencies $f_1$, $f_2$, and $f_1+f_2$, a peak will be generated in the principal domain of the bispectrum at the intersection between $f_1$ and $f_2$.
The bispectrum is the Fourier space (frequency domain) representation of the third-order cumulant of the time series of any dynamical quantity. Since the bispectrum is in the frequency domain, it is always a complex quantity with a magnitude and a phase. The input of the bispectrum is the time series and is calculated based on the following equation.
\begin{equation}\label{bispect_eqn}
    B(f_1,f_2) = \langle F(f_1) F(f_2) F^{\star}(f_1+f_2) \rangle
\end{equation}
Here, $\langle ... \rangle$ is the ensemble average over time windows, $F$ is the Fourier transform, and $F^\star$ is its complex conjugate. The $f_1$, $f_2$ are the two frequencies of the triad $(f_1, f_2, f_1+f_2)$. 
The bispectrum $B(f_1,f_2)$ is a function of two frequencies and is a non-zero quantity only if a phase coupling exists between the frequency triplet $f_1$, $f_2$, and $f_1+f_2$. The $B(f_1,f_2)$ is identically zero for spontaneously excited modes, \textit{i.e.,} the modes generated without phase coupling.  The normalized bispectrum of a time series yields the bicoherence and is given by~\cite{Rao_Gabr_SV_1984, Siu_IEEE_2008, Nosenko_PRE_2006, Ajaz_POP_2022}
\begin{equation}\label{bicoher_eqn}
    \gamma^2(f_1,f_2) = \frac{|B(f_1,f_2)|^2}{\langle |F(f_1)F(f_2)|^2 \rangle \langle |F^{\star}(f_1+f_2)|^2 \rangle} 
\end{equation}
\begin{figure*}[ht!]
    \centering
    \includegraphics[width=1.0\textwidth]{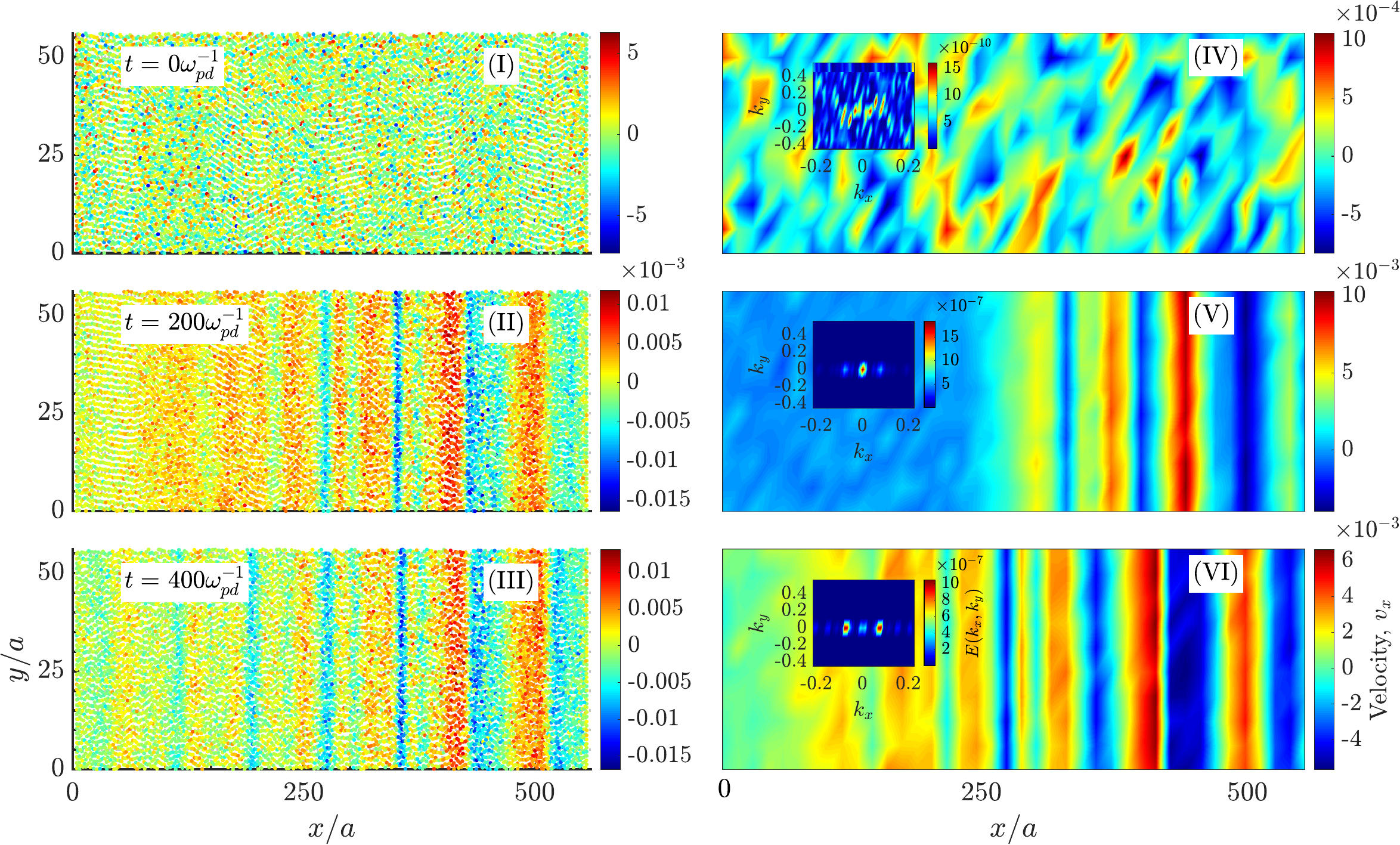}
    \caption{Time evolution of velocity perturbation excited by a single laser. Left column: velocity evolution in the xy domain. Right column: Fluidized velocity evolution in xy domain. Both show a clear indication of soliton trains. The inset in each subplot in the right column shows the evolution of the 2D energy spectrum $E(k_x, k_y)$.}
    \label{Fig_3}
\end{figure*}
Bicoherence quantifies phase coherence and is a measure of the fraction of power retained between the nonlinearly interacting modes. Ideally, the bicoherence is 1 for phase-coupled modes, \textit{i.e.,} modes generated due to coherent nonlinear interaction, and 0 for uncoupled modes, \textit{i.e.,} modes generated spontaneously. To obtain bicoherence, the total time series $T = M \times N$ is broken into N time series segments, each of M sampling data points. A correlation between coherent modes is statistically significant if the bicoherence condition $\gamma^2 > \sqrt{6/2N}$~\cite{Siu_IEEE_2008}.
\section{Comparison of NLM based on the Langevin simulation with the fKdV model}
\label{NLM_MD_fKdV}
\begin{figure*}
    \centering
    \includegraphics[width= \textwidth]{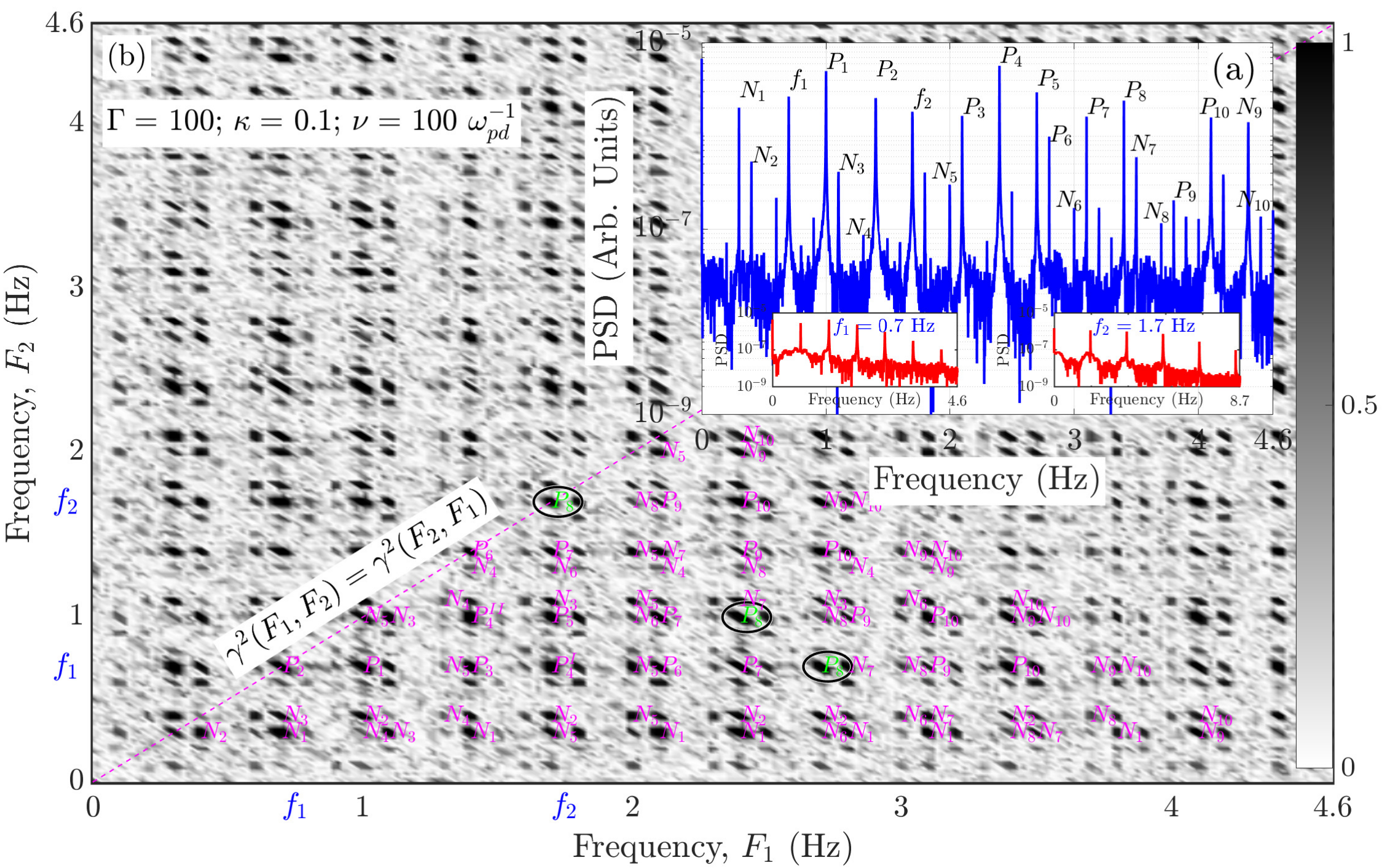}
    \caption{
    \textcolor{black}{NLM observed in YOCP using the Langevin MD simulations. 
     (a) PSD and (b) bicoherence of the time-series obtained from the time evolution of YOCP  for $\Gamma = 100$, $\kappa = 0.1$ and damping rate $\nu = 100\ \omega_{pd}^{-1}$. The small insets show the PSD of individual waves with frequency $f_1 = 0.7$ Hz and $f_2 = 1.7$ Hz. 
     }
     }
    \label{Fig_4}
\end{figure*}

To study the NLM process, two waves of frequency $f_1 = 0.7$ Hz and $f_2 = 1.7$ Hz were launched using the model forces given by Eqns.~\eqref{Eqn_WEF_L}, and~\eqref{Eqn_WEF_R} with $\Gamma = 100$ and $\kappa = 0.1$. The choice of frequencies was motivated by our earlier published work~\cite{Ajaz_POP_2020}, which showed excellent agreement with the dusty plasma experiment~\cite{Nosenko_PRL_2004}. 
We created three regions in the rectangular box. Two were used for wave excitation by the external forces, and one was used to collect the velocity time series in the region-of-interest (ROI) as shown in Fig.~\ref{Fig_1}. We first excited one wave for confirmation to see the nature and frequency of the wave by turning on only one external driving source and running the simulation for a sufficient time. The evolution of a single nonlinear wave in the system due to a single driving force \textcolor{black}{given by Eqn.~\eqref{Eqn_WEF_R}} is shown in Fig.~\ref{Fig_3}. The peak of the wave exciter is on the right side at $0.8L_x$; the wave fronts are seen to be propagating towards the left side. The left column (I-III) shows the propagation of the wave in time through the particle picture. The pseudo-color of particles represents their velocity amplitudes. The right column (IV-VI) shows the same evolution via a fluid picture obtained by binning particles over grids. The time evolution shows the propagation of waves from right to left with a damping due to the neutral drag $\bar{\nu}$. The amplitude is chosen such that the wave remains sufficiently nonlinear as it reaches the ROI. It is to ensure that nonlinear mixing occurs as two waves meet at the ROI. Insets show the wave numbers of excitation that display secondary harmonics at $t = 400~\omega_{pd}^{-1}$, establishing its nonlinear characteristics. 
\paragraph*{}
After confirming the frequency and nature of both waves individually, we turned on both sources to study NLM due to the interaction of the two waves. The time series of the system's evolution was collected in the ROI. Furthermore, the power spectrum density (PSD) and bi-coherence were calculated from the time series.
Figure~\ref{Fig_4} shows the PSD (a) and bicoherence (b) of NLM using Langevin MD simulations. The small insets show the nonlinear nature of two individual waves excited independently in the medium, as was explained in Fig.~\ref{Fig_2}. As can be seen, the nonlinear interaction of the two externally excited waves gives rise to a rich mixing spectrum. The inset (a) shows the PSD of the mixing profile. 
The fKdV model shows similar NLM profiles in the power spectrum and bicoherence with the same two driving frequencies: $f_1 = 0.7$ Hz and $f_2 = 1.7$ Hz~\cite{Ajaz_POP_2022}.
A comparison of the mixing profiles observed in the Langevin MD simulations and the solution of the fKdV model is shown in a tabular form in Table~\ref{Tab_2}.
In the fKdV model, the nonlinear interaction between the natural KdV mode and its harmonics with the external nonlinear driver gives rise to the mixing spectrum~\cite{Ajaz_POP_2022}.
\begin{table*}
\caption{Dominant modes observed in the Langevin MD simulations, and the fKdV model.}
\def~{\hphantom{0}}
{\renewcommand{\arraystretch}{1.5}
\begin{tabular}{|p{3.0cm}|p{2.0cm}|p{2.00cm}|p{3.0cm}|p{2.0cm}|p{2.00cm}|} 
\hline
Frequency (Hz) & fKdV model & {Langevin MD}   & 
Frequency (Hz) & fKdV model & {Langevin MD}  \\ 
\hline
 $f_1$                   & 0.7              & 0.7       
 &
 $N_1=f_2-2f_1$         & \checkmark       & \checkmark \\ \hline
 $f_2$                   & 1.7              & 1.7   
 &
 $N_2=3f_1-f_2$         & \checkmark       & \checkmark \\  \hline
 $P_1=f_2-f_1$           & \checkmark       & \checkmark  
 &
 $N_3=4f_1-f_2$         & \checkmark       & \checkmark \\  \hline
 $P_2=2f_1$              & \checkmark       & \checkmark  
 &
 $N_4=2f_2-3f_1$        & \checkmark       & \checkmark   \\  \hline
 $P_3=3f_1$              & \checkmark       & \checkmark 
 &
 $N_5=2(f_2-f_1)$       & \checkmark       & \checkmark  \\  \hline
 $P_4=f_1 +f_2$          & \checkmark       & \checkmark 
 &
 $N_6=3(f_2-f_1)$       & \checkmark       & \checkmark \\  \hline
 $P_5=2f_2-f_1$          & \checkmark       & \checkmark 
 &
 $N_7=5f_1$             & \checkmark       & \checkmark  \\  \hline
 $P_6=4f_1$              & \checkmark       & \checkmark 
 &
 $N_8=3f_2-2f_1$        & \checkmark       & \checkmark  \\ \hline 
 $P_7=2f_1+f_2$          & \checkmark       & \checkmark 
 &
 $N_9=3f_2-f_1$         & \checkmark       & \checkmark   \\  \hline
 $P_8=2f_2$              & \checkmark       & \checkmark  
 &
 $N_{10}=4f_1+f_2$      & \checkmark       & \checkmark  \\  \hline
 $P_9=3f_1+f_2$          & \checkmark       & \checkmark 
 &
                         &                  &             \\  \hline  
 $P_{10}=2f_2+f_1$       & \checkmark       & \checkmark   
 &
                        &                  &              \\   \hline
\end{tabular}
}
\label{Tab_2}
\end{table*}
The statistically significant bicoherence condition for the present simulations is found to be $\gamma^2 > \sqrt{6/2N} = 0.6124$, because we have divided the total time series $T = M \times N$ into N = 8 segments, each segment having M = 4096 sampling data points. The strength of modes occurring in the bicoherence plot (i.e., Fig.~\ref{Fig_4}), which have a magnitude much greater than $\gamma^2$, is statistically significant. Hence, they are retained and are designated to be originating from coherent nonlinear interactions. 
\paragraph*{}
The similarity in the spectral properties and nonlinear mixing characteristics is striking, given the many fundamental differences in the dynamical properties of the two models. The MD model tracks particle-level dynamics, while the fKdV model employs a fluid-based approach. In the MD model, viscosity naturally arises from particle interactions, whereas the fKdV model does not account for viscous damping. Furthermore, the MD model captures the full nonlinearity of the medium, whereas the fKdV model represents a weakly nonlinear realization of the full fluid-Poisson model for YOCP. 
Furthermore, the MD simulations have been carried out in two dimensions, while the fKdV model represents one-dimensional plasma dynamics. Moreover, there are intrinsic statistical fluctuations present in the Langevin simulations, which are inherently ignored by the fluid-based fKdV model. We are only interested in the dominant modes that have quantitatively equal strength both in the fKdV model as well as in the Langevin simulations of YOCPs. 
\paragraph*{}
A physical understanding of the similarity in the results can be partly gained from the fact that some of these differences might be minimized due to certain intrinsic dynamical conditions of the simulation. For example, the presence of frictional dissipation in the YOCP could reduce the amplitude of the primary waves to push the NLM process towards the weak coupling limit and thereby be in line with the fKdV approximation. 
They might also be responsible for the heating of the system due to energy transfer from the driving forces, thereby putting the YOCP system in a fluid state. \textcolor{black}{ The application of an external perturbation (laser force) leads to a rise in the system’s temperature, which in turn reduces the coupling strength. This reduction leads to a transition from a solid-like to a fluid-like regime.}
Likewise, the forces driving the two primary waves from the end walls have a one-dimensional nature (being independent of the y coordinate), which could be responsible for their close alignment with the fKdV model.
Moreover, since the fKdV explains the dynamics of 2D Yukawa lattices in a weakly nonlinear regime, it is expected that it can capture many features of the YOCPs. Our present work shows that this is indeed the case and that the fKdV does capture the essential features of the mixing profiles observed in the YOCP based on the Langevin simulations.
Perhaps the most notable result is the remarkable agreement in the power spectra and bicoherence spectra despite the inherent differences in the dynamics of the two systems and the differences in the manner of driving the systems. Thereby, our results broaden the scope of the applicability of the fKdV model for investigating NLM phenomena to a large class of systems that can be represented by the YOCP model. 
 \begin{figure*}
     \centering
     \includegraphics[width=\textwidth]{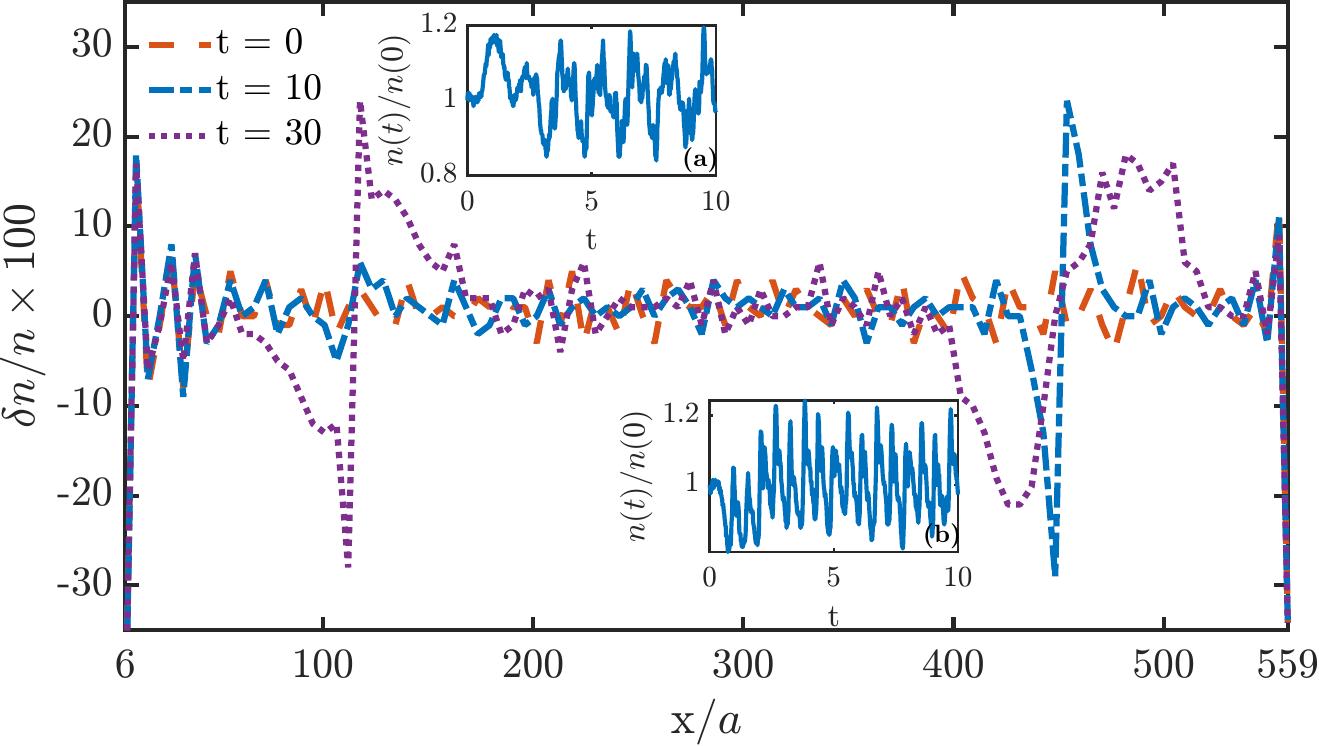}
     \caption{
     Density fluctuations at three different times due to external forcing. Inset (a) shows a time series collected from a region near the 0.7 Hz driving, while inset (b) shows a time series collected from a region near the 1.7 Hz driving. The density fluctuations reach up to $10-20 \%$ of the equilibrium density.}
     \label{Fig_5}
 \end{figure*}
\textcolor{black}{
This conclusion is further strengthened by the results displayed in Fig.~\ref{Fig_5} that highlights the density fluctuations observed in the present particle-based simulations. During the mixing process, density fluctuations of up to $10-20 \%$ of the equilibrium value have been recorded that follow the weak nonlinearity approximation of the KdV model.  The fact that the features of both the fKdV model and Langevin dynamics still closely follow the characteristic frequencies at moderate-amplitude fluctuations strongly suggests an extended validity of the fKdV model.
}
\section{Summary and conclusion}
\label{sum_con}
In this study, we have investigated nonlinear wave mixing in a driven YOCP system through first-principles Langevin MD simulations and compared our results with a similar study that was done using the fKdV model equation. We have found a remarkable similarity in the power spectra and the bicoherence spectra, indicating that the coherent three-wave coupling interaction dominates the NLM process.
Our results provide considerable strength to the applicability of the simple fKdV model to predict NLM phenomena in a broader class of systems that are represented by Yukawa interactions. The fKdV model equation is a well-established model that governs the nonlinear evolution of compressional waves in a 2D Yukawa lattice. The YOCP is valid for both weakly coupled 3D dusty plasma and 1D/2D dust crystal lattices. Therefore, fKdV is a good approximation that is valid for understanding the nonlinear mixing features observed because of the nonlinear interaction between the DAWs and the DLWs in a dusty plasma medium.
\paragraph*{}
Our study can be further extended to explore other nonlinear phenomena such as synchronization of waves in the presence of dissipation, as have been previously studied using a fKdV-Burger equation~\cite{Ajaz_PRE_2023} and thus provide a more realistic platform for comparison with past experimental results~\cite{Suranga_PRE_2012, Deka_PST_2020}.
\textcolor{black}{In the realm of synchronization, bispectral analysis can detect phase coupling between oscillatory components of interacting systems. Unlike linear coherence, which only identifies simple correlations, bispectral analysis reveals quadratic phase coupling, a key feature of nonlinear synchronization~\cite{Abe_SR_2024, Ajaz_PRE_2023}. 
In the field of neuroscience, studies have experimentally validated bispectral indicators of brain synchrony, highlighting their importance in both clinical applications and cognitive research~\cite{Ajaz_POP_2022, Siu_IEEE_2008, Kim_PRL_1997, Ziogas_IEEE_2023}.
In turbulence studies, this research plays a crucial role in identifying triadic interactions that drive energy cascades across different scales. For instance, in hydrodynamic turbulence, bispectral analysis has been instrumental in uncovering coherent structures and mode coupling, which are responsible for phenomena like intermittency and anomalous transport~\cite{Monsalve_PRL_2020, Kim_PRL_1997, Rao_Gabr_SV_1984, Lee_PRL_2016, Milligen_PRL_1995}. 
This approach not only enhances our understanding of complex nonlinear systems like plasmas but also establishes bispectral analysis as a vital tool in both theoretical and experimental domains.
}
\section{Acknowledgment}
This work was supported by the Indian Institute of Technology Jammu Seed Grant No. SG0012. 
AM and ST acknowledge the use of Agastya High-Performance Computing for the present studies. 
ST also acknowledges Science and Engineering Research Board (SERB) Grant No. CRG/2020/003653 for partial support for the work. FB thanks the University Grants Commission (UGC) India for the fellowship. AS is grateful to the Indian National Science Academy (INSA) for the INSA Honorary Scientist position. 
\bibliography{NLM_YOCP_fKdV}
\end{document}